\documentstyle[floats,aps,aipbook,graphics]{revtex}

\input psfig
\newdimen\mycirclesize
\newcount\mycirclesiz

\def\mycircl#1{%
\mycirclesiz\mycirclesize%
\mycirc{\the\mycirclesiz}%
}
\def\mycirc#1{%
\special{" newpath 0 0 #1 65536 div 0 360 arc stroke}%
}

\def\mydis#1#2{%
\mycirclesiz\mycirclesize%
\mydi{\the\mycirclesiz}{#2}%
}
\def\mydi#1#2{%
\special{" newpath 0 0 #1 65536 div 0 360 arc #2 setgray fill}%
}
\def\myoval#1#2{%
\mycirclesize #1\unitlength%
\myova{\mycirclesize}{#2}%
}
\def\myova#1#2{%
\mycirclesiz\mycirclesize%
\myov{\the\mycirclesiz}{#2}%
}
\def\myov#1#2{%
\special{" 1 #2 scale newpath 0 0 #1 65536 div 0 360 arc stroke}%
}
\begin{document}

\begin{titlepage}
\setcounter{footnote}{0}
\renewcommand{\thefootnote}{\fnsymbol{footnote}}
\begin{flushright}
CERN--TH/95--176\\
hep-ph/9506421\\
June 1995
\end{flushright}
\vspace{\fill}
\title{Jets in QCD\boldmath
\footnote{Talk given at the 10$^{\mathrm{th}}$ Topical Workshop on
Proton-Antiproton Collider Physics, Batavia, IL, May 9--13, 1995.}}
\author{Michael H. Seymour}
\address{Division TH, CERN, CH-1211 Geneva 23, Switzerland}
\maketitle
\thispagestyle{empty}
\begin{abstract}\pretolerance 1000
  Many analyses at the collider utilize the hadronic jets that are the
  footprints of QCD partons.  These are used both to study the QCD
  processes themselves and increasingly as tools to study other physics,
  for example top mass reconstruction.  However, jets are not
  fundamental degrees of freedom in the theory, so we need an {\em
    operational jet definition} and {\em reliable methods to calculate
    their properties}.  This talk covers both of these important areas
  of jet physics.
\end{abstract}
\vspace{\fill}
\par\noindent
CERN--TH/95--176\\
June 1995
\setcounter{footnote}{0}
\renewcommand{\thefootnote}{\alph{footnote}}
\end{titlepage}
$\mbox{}$\addtocounter{page}{-1}\thispagestyle{empty}\newpage

\title{Jets in QCD}
\author{Michael H. Seymour}
\address{Division TH, CERN, CH-1211 Geneva 23, Switzerland}
\maketitle
\begin{abstract}\pretolerance 1000
  Many analyses at the collider utilize the hadronic jets that are the
  footprints of QCD partons.  These are used both to study the QCD
  processes themselves and increasingly as tools to study other physics,
  for example top mass reconstruction.  However, jets are not
  fundamental degrees of freedom in the theory, so we need an {\em
    operational jet definition} and {\em reliable methods to calculate
    their properties}.  This talk covers both of these important areas
  of jet physics.
\end{abstract}

\section*{Introduction}
To fully analyze high-energy data, one would dearly love to measure the
distributions of final-state quarks and gluons.  However, owing to the
confinement of colour charge these are not the final-state particles of
the reaction, colourless hadrons are\footnote{To avoid confusion, I
  should point out at this stage that I use the word ``hadron'' rather
  loosely to mean any particle produced by the hadronization process,
  which also includes soft photons and leptons coming from secondary
  hadron decays.}.  This means that we should instead discuss {\em event
  properties\/} that satisfy the following conditions:
\begin{itemize}
\item Well-defined and easy to measure from the hadronic final-state.
\item Well-defined and easy to calculate order-by-order in perturbation
  theory from the partonic final-state.
\item Have a close correspondence with the distributions of the
  final-state quarks and gluons that we are really interested in.
\end{itemize}
These event properties are generally called {\em jets}.

It should be immediately apparent that there will be many event
properties that satisfy these conditions and hence many possible ways of
defining jets.  Although there are certainly some that are better than
others, in the sense that they are more reliably measurable or
calculable, we take the view for now that all definitions are created
equal.  There is of course an important corollary to this~--- since the
definition is not unique, it is not surprising to find that results
depend on it: What You See Is What You Look For, or
WYSIWYLF \cite{wysiwylf}.

The last of these three conditions is the cause of a great deal of
confusion, because it has the direct practical consequence that in
leading order perturbative calculations there is a one-to-one
correspondence between partons and jets.  This can make it very easy to
make the mistake of thinking that we are actually measuring a primary
parton, instead of merely an event property that is largely determined
by a primary parton.  It also means that all jet definitions give equal
results at leading order and it is only at higher orders that one can
calculate the dependence on the definition, which is certainly seen in
the data \cite{results}.  This underlines the importance of making jet
calculations beyond leading order, either exactly in perturbation theory
as described in Walter Giele's talk \cite{walter}, or using parton shower
methods as I describe below.

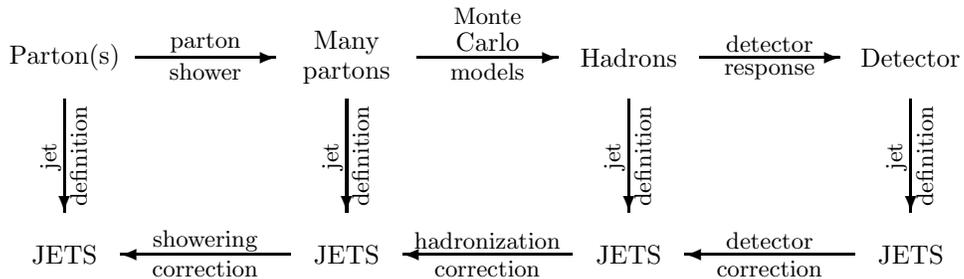
\begin{figure}[b]
\vspace*{-3ex}
  \begin{center}
    \unitlength 0.75cm
    \begin{picture}(15,5)(1.5,-0.5)
      \thicklines
      \put(1.5,4){\makebox(0,0){Parton(s)}}
      \put(6.5,4){\makebox(0,0){\shortstack{Many\\partons}}}
      \put(11.5,4){\makebox(0,0){Hadrons}}
      \put(16.5,4){\makebox(0,0){Detector}}
      \put(1.5,3.25){\vector(0,-1){2}}
      \put(6.5,3.25){\vector(0,-1){2}}
      \put(11.5,3.25){\vector(0,-1){2}}
      \put(16.5,3.25){\vector(0,-1){2}}
      \put(2.75,4){\vector(1,0){2.5}}
      \put(7.75,4){\vector(1,0){2.5}}
      \put(12.75,4){\vector(1,0){2.5}}
      \put(1.5,0.5){\makebox(0,0){JETS}}
      \put(6.5,0.5){\makebox(0,0){JETS}}
      \put(11.5,0.5){\makebox(0,0){JETS}}
      \put(16.5,0.5){\makebox(0,0){JETS}}
      \put(5.5,0.5){\vector(-1,0){3}}
      \put(10.5,0.5){\vector(-1,0){3}}
      \put(15.5,0.5){\vector(-1,0){3}}
      \put(4,4.25){\makebox(0,0){\small parton}}
      \put(4,3.75){\makebox(0,0){\small shower}}
      \put(9,4.5){\makebox(0,0){\shortstack{\small Monte\\Carlo}}}
      \put(9,3.75){\makebox(0,0){\small models}}
      \put(14,4.25){\makebox(0,0){\small detector}}
      \put(14,3.75){\makebox(0,0){\small response}}
      \put(4,0.75){\makebox(0,0){\small showering}}
      \put(4,0.25){\makebox(0,0){\small correction}}
      \put(9,0.75){\makebox(0,0){\small hadronization}}
      \put(9,0.25){\makebox(0,0){\small correction}}
      \put(14,0.75){\makebox(0,0){\small detector}}
      \put(14,0.25){\makebox(0,0){\small correction}}
      \put(1.25,2.25){\rotatebox{90}{\makebox(0,0){\small jet}}}
      \put(1.75,2.25){\rotatebox{90}{\makebox(0,0){\small definition}}}
      \put(6.25,2.25){\rotatebox{90}{\makebox(0,0){\small jet}}}
      \put(6.75,2.25){\rotatebox{90}{\makebox(0,0){\small definition}}}
      \put(11.25,2.25){\rotatebox{90}{\makebox(0,0){\small jet}}}
      \put(11.75,2.25){\rotatebox{90}{\makebox(0,0){\small definition}}}
      \put(16.25,2.25){\rotatebox{90}{\makebox(0,0){\small jet}}}
      \put(16.75,2.25){\rotatebox{90}{\makebox(0,0){\small definition}}}
    \end{picture}
  \end{center}
\vspace*{-3ex}
\caption[]{Schematic diagram of a typical analysis using jets.}
\label{typical}
\end{figure}

A typical analysis proceeds according to the general pattern shown in
Fig.~\ref{typical}.
In QCD studies, the individual processes that take us from left to right
are interesting in their own right, while for other studies the really
crucial question is how well these processes are modeled so that
parton-level predictions can be compared with detector-level
measurements.

There are several important issues that need to be addressed when making
such analyses:
\begin{itemize}
\item Which of the four things labeled ``jets'' should we be aiming to
  measure and calculate?
\item How well do we understand this loop?
\item How can we improve our understanding and predictions?
\end{itemize}
The first of these is a question of demarcation~--- should theorists be
correcting their predictions to hadron level, or should experimenters be
correcting their data to parton level and is largely a question of
personal taste.  The other two questions are the main themes of this
talk.

I will begin by describing the current norm in jet definitions~--- the
cone algorithm.  In fact I should say cone algorithm{\Large\bf s}, since
everyone seems to have their own slightly different preferred version.
Then I will describe a recently proposed family of cluster-based
`$k_\perp$' jet algorithms, which are descended from those used in
$\mathrm{e^+e^-}$ annihilation, and which have a variety of advantages
over cone-based algorithms.

Besides the exact matrix-element calculations described in Walter's
talk, the main method used to calculate jet properties is the parton
shower approach, either implemented as Monte Carlo simulations, or
\nopagebreak[3]
explicitly as analytical calculations.  I will describe the common
theoretical basis of these, as well as a few more technical points
associated with the Monte Carlo algorithms.  I will also highlight an
important area in which improvements can be expected in the near future.

Hadronization corrections are not understood at a fundamental level at
present and are generally estimated using Monte Carlo programs.
However, since they do not play too big a r\^ole in most collider jet
studies I do not spend too much time describing them.

\vspace*{-1ex}
\section*{Jet Definitions}
\vspace*{-1ex}
The first two requirements of a jet definition, that the jets be easy to
measure and calculate reliably, turn out to be very similar.  This has
meant that experimental and theoretical improvements have tended to go
hand in hand, with modifications proposed on theoretical grounds tending
to result in improvements in the experimental properties of the
algorithm and vice versa.  This is because many of the problems are in
fact the same.

In perturbative QCD there is a collinear divergence when any two
massless partons are parallel.  In the {\em total\/} cross-section, this
divergence is guaranteed to be canceled by a contribution from the
virtual correction to the equivalent process, with the two partons
replaced by their sum.  However, for this cancelation to also take place
in the jet calculation, it is necessary to ensure that a collinear pair
of particles are treated identically to a single particle with their
combined momentum.  This means that algorithms that use information such
as which particle in the event had the highest energy cannot be
calculated perturbatively, since the resulting jet properties could be
altered by replacing the hardest particle in the event by two or more
collinear particles none of which is any longer the hardest one in the
event.  From the experimental point of view, the equivalent problem is
the fact that parallel particles go into the same calorimeter cell and
can never be resolved.  Thus any algorithm that depended critically on
resolving a pair of almost collinear particles would give results that
depended strongly on the angular resolution of the calorimeter.

Likewise the requirement of infrared safety, i.e.~insensitivity to
emission of low energy particles, is necessary in perturbative
calculations to avoid the soft divergence and in experiments to avoid
bias from the threshold trigger of a calorimeter cell and the background
noise.

Another requirement that has recently become topical in
$\mathrm{e^+e^-}$ and DIS physics is that the definition be fairly local
in angle.  This is experimentally useful because of the transverse size
of a hadronic shower in the calorimeter~--- often the energy from a
single hadron is spread over several calorimeter towers, and it is clear
that the jet definition should tend to put all this energy into the same
jet.  However, with the JADE algorithm, which has been the standard in
those collisions for some time, this is not always the case.  Also, if
the hadron happens to be near the edge of the jet cone, a cone algorithm
will neglect the energy falling outside it.  On the other hand, in the
$k_\perp$ cluster algorithm, all this energy will be clustered together,
before the final decision is made about whether to include it into the
\nopagebreak[3]
jet is taken.  This also results in improved theoretical properties as I
will describe below.

\subsection*{Cone Algorithms}
Cone algorithms have been the standard way to define jets in $\bar pp$
collisions for many years.  They are conceptually very simple to define,
as a {\em direction that maximizes the energy flowing through a cone
  drawn around it}.  However, the complications start when you consider
what happens when two of these cones overlap.

Although the `Snowmass Accord' \cite{snowmass} agreed in principle the
details of the cone algorithm, so that theorists and experimenters could
use the same definition, this very important issue was neglected and has
plagued jet studies ever since.  It has been found that the properties
of the resulting jets depend strongly on the exact treatment of the
overlap region.  This is not really a problem, since we do expect that
different jet definitions will give different jets.  The big problem is
the fact that the way the properties change depend on the number of
particles in the jet, giving very big differences between the jet
properties at parton, hadron and detector level.

\begin{figure}[b]
\vspace*{-3ex}
  \begin{center}
    \unitlength 0.75cm
    \begin{picture}(15,5)
      \put(2.5,0.5){\line(0,1){2}}
      \put(2.2,2){\makebox(0,0){$R_1$}}
      \put(2.8,2){\makebox(0,0){$R_2$}}
      \thicklines
      \put(2.5,0.5){\line(1,2){2}}
      \put(2.5,0.5){\line(-1,2){2}}
      \put(2.5,4.5){\myoval{2.5}{0.2}}
      \put(10.5,0.5){\line(1,4){0.7}}
      \put(10.5,0.5){\line(1,3){1.07}}
      \put(10.5,0.5){\line(2,5){1.44}}
      \put(10.5,0.5){\line(1,2){2}}
      \put(10.5,0.5){\line(3,5){2.16}}
      \put(10.5,0.5){\line(2,3){2.13}}
      \put(10.5,0.5){\line(3,4){2.1}}
      \put(10.5,0.5){\line(-1,4){0.7}}
      \put(10.5,0.5){\line(-1,3){1.07}}
      \put(10.5,0.5){\line(-2,5){1.44}}
      \put(10.5,0.5){\line(-1,2){2}}
      \put(10.5,0.5){\line(-3,5){2.16}}
      \put(10.5,0.5){\line(-2,3){2.13}}
      \put(10.5,0.5){\line(-3,4){2.1}}
      \put(8.5,4.5){\myoval{2.5}{0.2}}
      \put(12.5,4.5){\myoval{2.5}{0.2}}
    \end{picture}
  \end{center}
\vspace*{-3ex}
\caption[]{Schematic diagram of a jet configuration in which the
  cone algorithm at parton- and hadron-level give very different
  answers.}
\label{camel}
\end{figure}
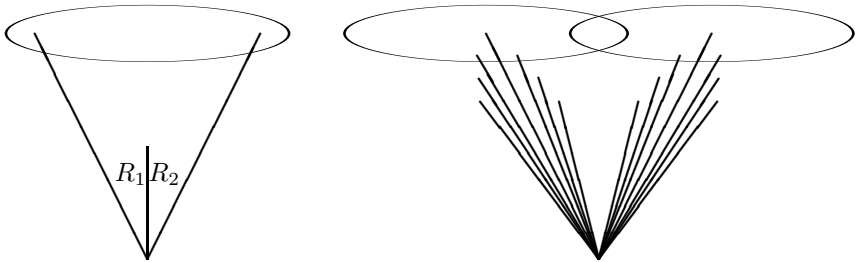

\begin{figure}
\centerline{\psfig{file=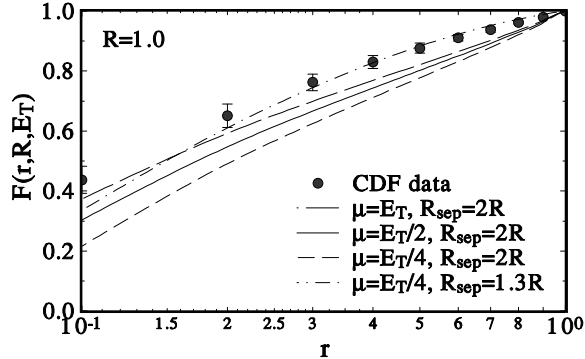,height=2in}}
\caption[]{The energy profile of a 100~GeV jet, taken from~\cite{camel}.}
\label{CDFprofile}
\end{figure}

One example of this effect was studied in detail by Steve Ellis and
company \cite{camel} when comparing their parton-level predictions with
CDF data.  They found that owing to the width of a hadronic jet,
configurations that were called one jet in their calculation could be
called two jets at hadron level, as illustrated in Fig.~\ref{camel}.
They found that the hadron-level algorithm could be simulated at parton
level by introducing a parameter $R_{\mathrm{sep}},$ in addition to the
jet radius $R,$ such that two partons are merged into one jet only if
the conditions
\[
  R_1<R \quad\&\quad R_2<R \quad\&\quad R_1+R_2<R_{\mathrm{sep}}
\]
are satisfied.  The first two are the Snowmass Accord, while the third
is the additional requirement that the partons not be too close to the
edge of the cone.  It was found that good agreement with CDF
data \cite{CDFprofile} could be obtained by setting
$R_{\mathrm{sep}}=1.3R,$ as shown in Fig.~\ref{CDFprofile}.
However, the danger in this is that the value of $R_{\mathrm{sep}}$ is
not given to us by the jet definition and could be viewed as a
phenomenological parameter that should be tuned to data.  In this sense,
there is no reason to suppose that the value of $1.3R$ should also
describe jets in the far forward region, or jets produced in a
completely different reaction, like top quark decays.

One very nice feature of cone algorithms is the apparent ease with which
energy corrections can be made.  This is because they are purely
geometrical, so the amount of out-of-cone showering in the calorimeter
for example, can be very easily calculated from the known detector
response and the energy inside the cone.  Another type of correction
that is sometimes applied is for the amount of the jet's energy that is
radiated outside the cone and the amount of energy from the underlying
event that sneaks into the cone.  When comparing with leading order
calculations this may appear straightforward since they do not account
for the energy spread of the jet.  However, at next-to-leading order,
the calculation already includes the fact that some of a jet's energy is
radiated outside the cone, so this should not be corrected for.
Furthermore, the underlying event correction is inevitably
model-dependent, even when it is measured from data.  This is because in
some models the jet pedestal is correlated with the jet momentum and
direction, while in others it is completely uncorrelated.  Thus
different experimental procedures such as subtracting the average seen
in a cone in minimum bias events or the average seen in a random
direction in jet events at the same jet transverse momentum will give
results that are interpreted differently in different models.  The
effect of such assumptions on jet data is discussed in detail
in \cite{steve}.  An experimental test was proposed in \cite{pedestal}
that would shed a lot of light on these problems, since it claims to
disentangle how correlated the jet activity and pedestal height are.

\subsection*{\boldmath$k_\perp$ Clustering Algorithms}
Clustering algorithms have been the main way of defining jets in
$\mathrm{e^+e^-}$ annihilation for many years.  They work in a very
different way from cone algorithms~--- instead of globally finding the
jet direction, they start by finding pairs of particles that are
`nearby' in phase-space and merging them together to form new
pseudoparticles.  This continues iteratively until the event consists of
a few well-separated pseudoparticles, which are the output jets.  There
is however, no unique definition of `closeness' in phase-space and
different definitions define different algorithms.  Traditionally
invariant mass was used, but this has the disadvantage that it is
extremely non-local in angle~--- if two particles have low enough
energy, they will be merged together, regardless of how far apart they
are in angle \cite{nick}.  This can be solved by using as closeness the
momentum of the softer particle transverse to the axis of the
harder \cite{durham}, as in the $k_\perp$ jet algorithm\footnote{The
  $k_\perp$ jet algorithm for $\mathrm{e^+e^-}$ is sometimes called the
  `Durham algorithm'.}.  This means that, roughly speaking, the
algorithm repeatedly merges the softest particle in the event with its
nearest neighbour in angle.  It has the great theoretical advantage that
it allows the phase-space for multiple emission to be factorized in the
same way as the QCD matrix elements, allowing analytical parton shower
techniques to be used \cite{cdfw}, as I describe below.

However, in collisions with incoming hadrons, there are additional
particles in the final state that are not associated with hard jets,
namely those coming from the hadron remnant and underlying event.  For
many years this was seen as a barrier to using clustering algorithms in
hadronic collisions, as their property of exhaustively assigning every
final state particle to a jet is clearly unphysical there.  This problem
was considered in~\cite{cdw}, where it was shown that it could be
overcome by introducing an additional particle into the event by hand,
parallel to the incoming beams.  In the case of the $k_\perp$ algorithm,
this extra particle can be considered as having infinite momentum.  The
resulting jet cross-sections are guaranteed to satisfy the factorization
theorem, so absolute predictions using p.d.f.s measured in other
processes can be made for their values, unlike earlier attempts to solve
this problem.

As far as the practical properties of the algorithm are concerned, it is
essential for jet algorithms for hadron-hadron collisions to be
invariant under longitudinal boosts along the beam direction.  A set of
longitudinally-invariant $k_\perp$-clustering algorithms for
hadron-hadron collisions was proposed in~\cite{cdsw}.  Briefly, the
algorithm proceeds as follows:
\begin{enumerate}
\item
  For every pair of particles, define a closeness
\[
  d_{ij} = \min(E_{Ti},E_{Tj})^2\Delta R^2, \quad\quad
  \Delta R^2 \equiv \Delta\eta^2+\Delta\phi^2.
\]
  Note that for small opening angle, $\Delta R\ll1,$ we have
\[
  \min(E_{Ti},E_{Tj})^2\Delta R^2 \approx
  \min(E_i,E_j)^2\Delta\theta^2 \approx k_\perp^2.
\]
\item
  For every particle, define a closeness to the beam particles,
\[
  d_{ib} = E_{Ti}^2 R^2,
\]
  where $R$ is an adjustable parameter of the algorithm introduced
  in~\cite{es}.
\item
  If $\min\{d_{ij}\}<\min\{d_{ib}\},$ {\em merge\/} particles $i$ and
  $j$.
\item
  If $\min\{d_{ib}\}<\min\{d_{ij}\},$ jet $i$ is {\em complete}.
\end{enumerate}
These steps are iterated until a given {\em stopping condition\/} is
satisfied, as I discuss in more detail in a moment.  Different ways of
merging two four-vectors into one define different schemes~--- the two
most common are the ``$E$''-scheme (simple four-vector addition) and the
``$p_t$''-scheme:
\begin{eqnarray*}
  E_{Tij} &=& E_{Ti} + E_{Tj}, \\
  \eta_{ij} &=& \left(E_{Ti}\eta_i+E_{Tj}\eta_j\right)/E_{Tij}, \\
  \phi_{ij} &=& \left(E_{Ti}\phi_i+E_{Tj}\phi_j\right)/E_{Tij},
\end{eqnarray*}
which become equivalent for small opening angle.  Although there are
some small practical differences, they are not important here.

\begin{figure}[b]
\vspace*{-1ex}
\centerline{\psfig{file=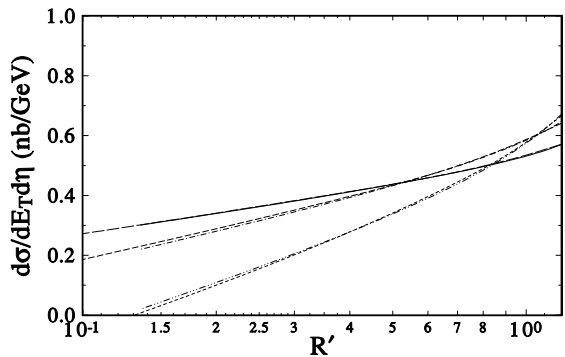,height=2in}}
\caption[]{The inclusive jet cross-section for central jets with
  $E_T=100$~GeV, as a function of $R$ for the cluster algorithm and
  $1.35R_{\mathrm{cone}}$ for the cone algorithm, for three different
  values of the renormalization and factorization scales, $\mu$.  The
  barely discernible pairs of curves are from the cone and cluster
  algorithms.  Taken from~\cite{es}, where more precise details of the
  calculation can be found.}
\label{nice}
\vspace*{-1ex}
\end{figure}

Depending on what kind of studies one is interested in, different
stopping conditions are useful.  For {\em inclusive} jet studies, one
iterates the above steps until all jets are complete \cite{es}.  In this
case, all opening angles within each jet are $<R$ and all opening angles
between jets are $>R$.  This means that the resulting jets are very
similar to those produced by the cone algorithm, with $R\sim
R_{\mathrm{sep}}$.  As shown in Fig.~\ref{nice}, this is certainly true
of the inclusive jet cross-section, for which the two algorithms are
almost indistinguishable at next-to-leading order.
It is worth noting that for a relatively soft jet, $d_{ij}$ is its
transverse momentum relative to the nearest hard jet, while $d_{ib}$ is
$R^2$ times its transverse momentum relative to the incoming beams.  If
$d_{ib}$ is the smaller then the jet is treated as initial-state
radiation, giving a resolvable jet, while if it is the larger it is
treated as final-state radiation, being merged into the other jet.  Thus
the value $R=1$ is strongly preferred theoretically, as it treats
initial- and final-state radiation on equal footings and would be
expected to give smallest higher-order corrections, at least from the
dominant logarithmic regions of phase-space.  In this sense, the
empirical relationship between $R$ and $R_{\mathrm{cone}}$ can be used
as an {\em a posteriori\/} justification for using
$R_{\mathrm{cone}}=0.7$.  The first experimental results using this
algorithm were reported in \cite{kate}.

In many cases, one instead wants to reconstruct exclusive final states,
for example in top quark decay.  In this case, one iterates the above
steps until all jet pairs have $d_{ij}>d_{\mathrm{cut}},$ an adjustable
parameter of the algorithm.  All complete jets with
$d_{ib}<d_{\mathrm{cut}}$ are discarded (merged with the beam remnants).
Thus $d_{\mathrm{cut}}$ acts as a global cutoff on the resolvability of
emission, initial- or final-state.  In this case setting $R=1$ is even
more strongly recommended, as in the original algorithm of~\cite{cdsw}.
One can either fix $d_{\mathrm{cut}}$ {\em a priori,} only resolving
radiation above the given $k_\perp$ cutoff, or adjust it event-by-event
to reconstruct a given number of final-state jets.

\begin{figure}[b]
\vspace*{-4ex}
\centerline{\psfig{file=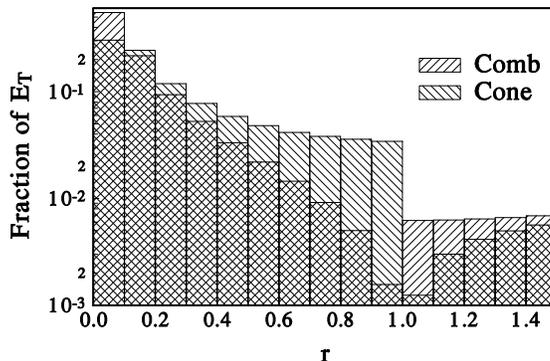,height=2in}}
\vspace*{-1ex}
\caption[]{The energy profiles of 100 GeV cone and cluster jets, taken
  from~\cite{es}.  The cluster jets (labeled ``Comb'') have more energy
  concentrated in the centremost bin, while the cone jets are more
  diffuse.  At the jet edge, $r=R=R_{\mathrm{cone}}=1,$ the trend is
  reversed~--- the cluster algorithm ignores energy more than $R$ away
  so gives a flat background contribution, while the cone algorithm has
  tried to pull this energy into the jet, leaving a deficit outside.}
\vspace*{-1ex}
\label{profile}
\end{figure}

\begin{figure}[b]
\vspace*{-3ex}
\centerline{\psfig{file=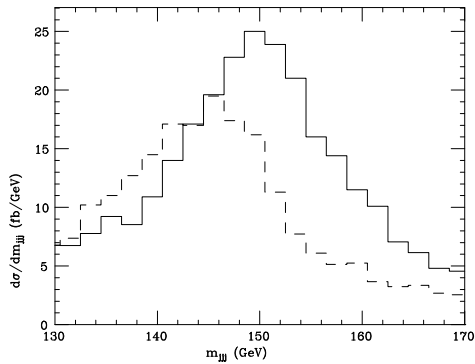,height=2in}}
\caption[]{Reconstructed three-jet mass distribution of top quark
  candidates, according to the cluster (solid) and cone (dashed)
  algorithms, at calorimeter level, for a top quark of nominal mass
  150~GeV.  Taken from~\cite{comparison}.}
\vspace*{-1ex}
\label{comparison}
\end{figure}

Although the cluster jets are very similar to the cone jets in terms of
the inclusive cross-section, they have several practical advantages.
Firstly, the jet overlap problem has completely disappeared~--- the
algorithm unambiguously assigns every particle to a single jet.  It does
this in a dynamic way, adjusting to the shapes of individual jets, and
so performs better than any fixed strategy such as drawing a dividing
line half way between the centres or giving all the energy to the
higher-energy of the two.  One can of course always come up with
pathological jet configurations where any given algorithm does not
cluster in the way that seems natural, but the relative contribution of
such configurations is much smaller for the cluster algorithm than the
cone.  This means that exactly the same algorithm can be used on
hadronic final states with many particles as on partonic final states
with one or two, without the need for additional adjustable parameters.
Secondly, the cluster algorithm is much less sensitive to perturbations
from soft particles than the cone algorithm, which results in smaller
hadronization and detector corrections, as well as a reduction in the
model-dependence of these corrections.  This is because in some sense it
pays the most attention to the core of the jet and only merges other
neighbouring particles if they are near enough to do so, whereas the
cone algorithm, which seeks to maximize the jet energy, does its best to
pull in as much neighbouring energy as possible, as illustrated in
Fig.~\ref{profile}.
This feature means that the $k_\perp$ cluster algorithm is particularly
suited for kinematic reconstruction of particle decays.  A detailed
Monte Carlo comparison was made in~\cite{comparison} for the cases of
top quark reconstruction at Tevatron energy and Higgs boson at LHC
energy.  The results for the former are shown in Fig.~\ref{comparison}
for the mass of the three-jet combination that minimizes
\[
  \chi^2 \equiv
  \left( \frac{m_{\mathrm{jj}}-\langle m_{\mathrm{jj}}\rangle}
    {\mbox{5 GeV}} \right)^2 +
  \left( \frac{m_{\mathrm{jjj}}-m_{\ell\nu\mathrm{j}}}
    {\mbox{10 GeV}} \right)^2,
\]
after the cut $\chi^2<3$.  The other cuts are described
in~\cite{comparison}.  The cluster algorithm's mass distribution is
centred on the true value and is much more symmetric than the cone
algorithm's.  Also the efficiency of the cluster algorithm is
considerably higher, owing to the greater cleanliness of the event
reconstruction, so that even though the widths of the two distributions
are similar, the cluster algorithm would give a smaller error on the
reconstructed mass.  Another reason for this greater cleanliness is that
the cluster algorithm is somewhat like a cone algorithm with its radius
adjusted event-by-event to suit the individual event dynamics~--- when
there are two hard jets near each other the effective radius is small to
resolve them well, when they are all far apart it is large, allowing a
good measurement of their energies.

\subsection*{Internal Jet Structure}
It is clearly important to study the internal structure of jets, both as
a test of QCD and to understand the efficiencies, corrections, etc, of
jet reconstruction for other studies.  In the cone algorithm, the
natural way to do this is by studying the spread of energy over various
annular radii, as shown in Figs.~\ref{CDFprofile} and~\ref{profile}.  In
the cluster algorithm, one has the possibility to study the internal
structure in a way that is much more like how we believe jets
develop~--- their structure comes not from a general smearing in angle,
but by radiating individual partons that give rise to smaller `sub-jets'
within the jet.  Thus by resolving these sub-jets we can compare their
distributions with those predicted by QCD or by Monte Carlo models.

Sub-jets are defined by first running the inclusive version of the
algorithm to find a jet.  Then the algorithm is rerun starting from only
the particles that were part of that jet, stopping when all pairs have
\[
  d_{ij} > y_{\mathrm{cut}}E_T^2,
\]
where $E_T$ is the transverse momentum of the jet and $y_{\mathrm{cut}}$
is an adjustable resolution parameter.  For $y_{\mathrm{cut}}\sim1$ the
jet always consists of only 1 sub-jet and for $y_{\mathrm{cut}}\to0$
every hadron is considered a separate sub-jet, so adjusting
$y_{\mathrm{cut}}$ allows us to go smoothly into the hadronization
region in a well-defined way and to study the parton$\to$many
partons$\to$hadrons transition in great detail.  Further details and the
first experimental results can be found in~\cite{rich}.  Similar studies
have also been performed in $\mathrm{e^+e^-}$
annihilation \cite{opalsub,alephsub} and will allow direct comparisons of
the two sources of jets.

\begin{figure}[h]
\vspace*{-3ex}
\centerline{\psfig{file=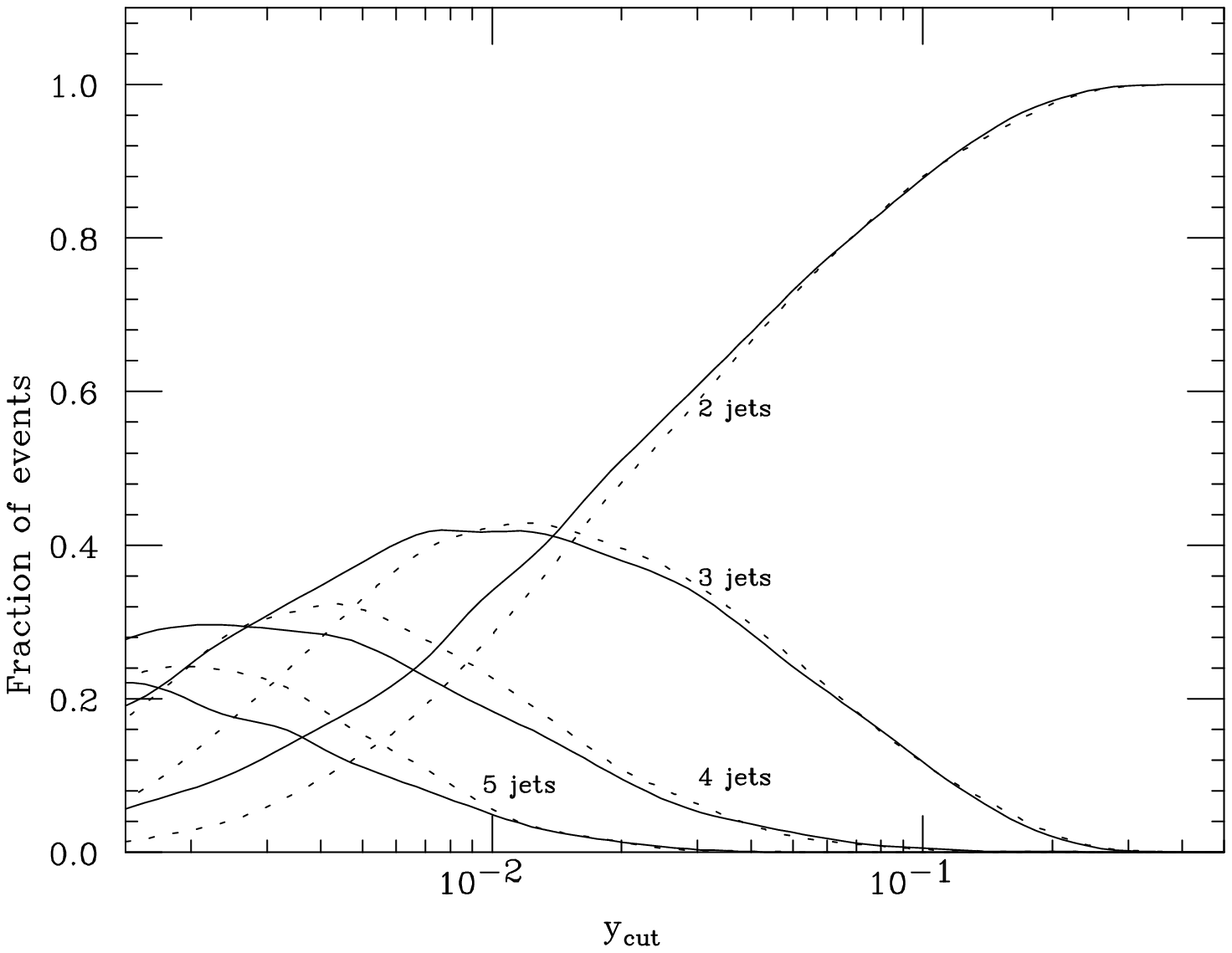,height=1.8in}%
\psfig{file=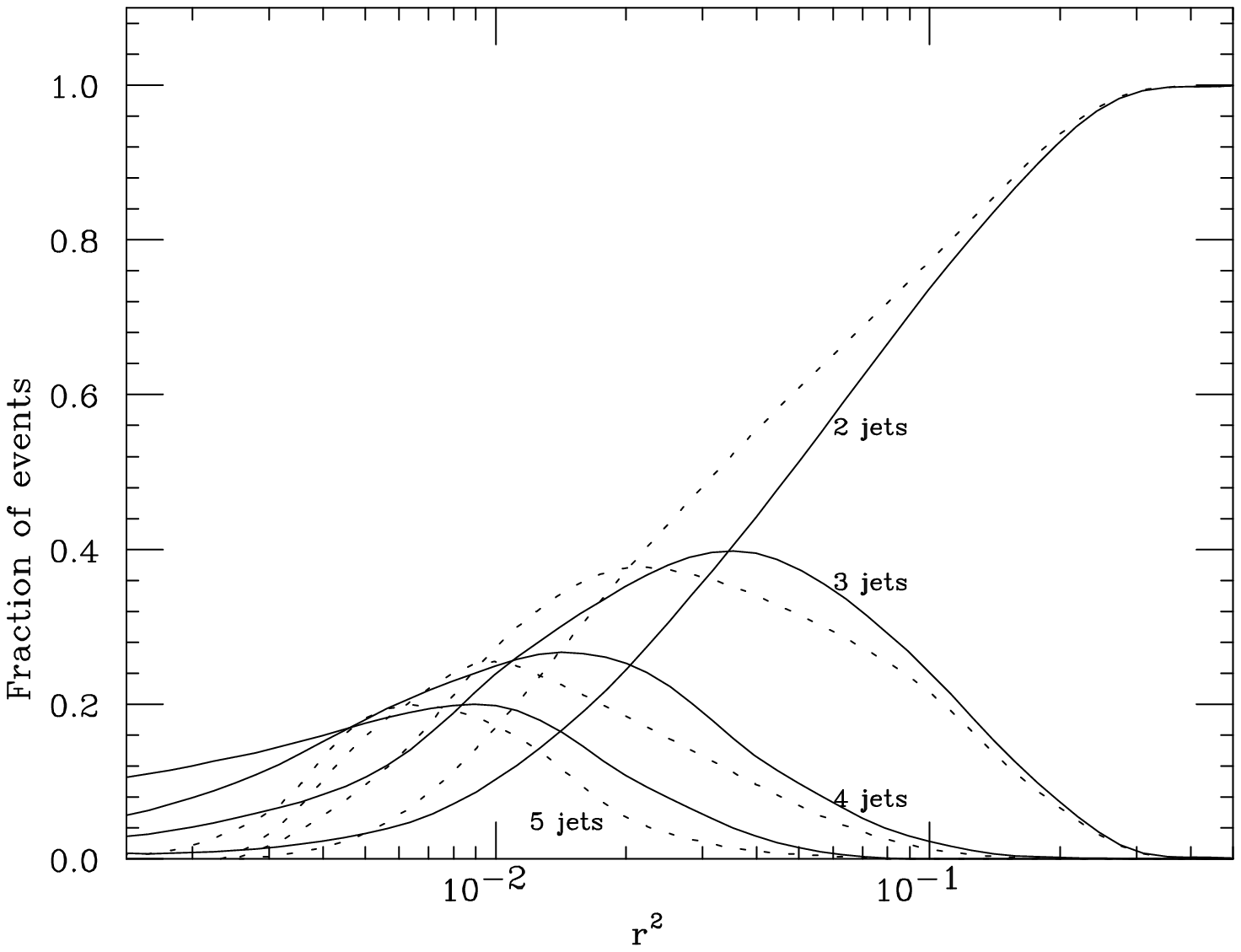,height=1.8in}}
\caption[]{The fraction of two-jet events that contain 2, 3, 4, or 5
  sub-jets when resolved at a scale $y_{\mathrm{cut}}$ for the cluster
  algorithm and a radius $r$ for the cone algorithm, at the parton level
  (solid) and calorimeter level (dotted).  Taken from~\cite{moriond}.}
\label{sub-jets}
\vspace*{-3ex}
\end{figure}

Of course, sub-jets can also be defined in the cone algorithm, by
rerunning using a smaller jet radius.  Results of a Monte Carlo
comparison are shown in Fig.~\ref{sub-jets}, where it can be seen that
the cone algorithm has much larger hadronization correction.
The reasons for this are discussed in~\cite{moriond}\footnote{Note that
  this figure supersedes the one in~\cite{cdsw} because that used a cone
  algorithm that was not infrared safe, so it is not surprising that it
  performed badly.  On the contrary, the one shown here uses the cone
  algorithm recommended by the authors of~\cite{camel}.  The event
  definition is also somewhat different to the one described above, but
  the conclusions would not be expected to be affected by this.}.

\section*{Calculating Jet Properties}
To study QCD, and to understand the corrections and efficiencies of jet
reconstruction, we are particularly interested in the {\em internal\/}
\nopagebreak[3]
properties of jets.  We have already seen in Walter's talk \cite{walter}
that external jet properties such as the transverse momentum spectra,
rapidity distributions, dijet mass spectra, etc, are well-described by
next-to-leading order QCD calculations.  The internal properties that I
will be concerned with are things like the energy profile of a jet, jet
mass distributions and internal sub-jet structure, which are
infrared-finite and can be calculated in perturbation theory.  However,
owing to the one-to-one correspondence between partons and jets in
leading order calculations, the leading non-trivial term is the one
given by ``next-to-leading'' order jet calculations.  This means that
they suffer all the usual problems with leading order calculations, like
large sensitivity to the renormalization scale.

For many of these jet properties, one encounters large logarithmic terms
at all orders in perturbation theory and it is essential that these
terms are reorganized (``resummed'') into an improved perturbation
series.  Examples include the energy profile at small angles and the
sub-jet structure at small $y_{\mathrm{cut}}$.  This is conventionally
done by the parton shower approach, either as a Monte Carlo simulation,
or analytically.

\vspace*{-1ex}
\subsection*{Parton Showers}
\vspace*{-1ex}
The cross-section for multiple emission factorizes in the collinear
limit.  This means that the production of additional partons close to a
jet direction can be described in a time-ordered probabilistic way as a
series of $1\to2$ splittings one after another.  In the strongly-ordered
limit, in which each splitting is much more collinear than the last,
this is guaranteed to reproduce the exact multi-parton matrix element.
This strongly-ordered region of phase-space contributes the leading
logarithmic contribution to energy-weighted quantities like the energy
profile.  Thus any parton shower based on sequential collinear emission
can predict these properties to leading logarithmic accuracy.  It is
worth noting that in the strongly-ordered limit, all measures of
collinearity, like transverse momentum, virtuality and angle, are
completely equivalent~--- algorithms that use different choices differ
only by next-to-leading logarithms.

\begin{figure}
\centerline{\psfig{file=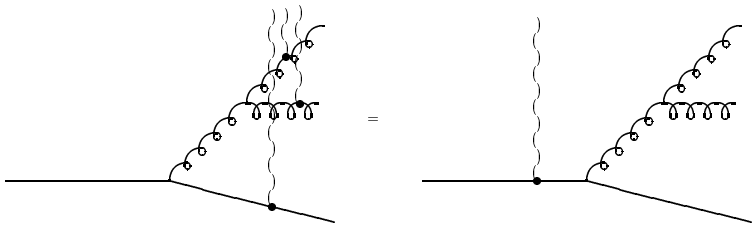,height=1.3in}}
\caption[]{Soft large-angle gluons cannot resolve the individual colour
  charges in the jet, so the coherent sum of emission off all the
  external lines is equivalent to emission by the original parton,
  imagined to be on-shell, i.e.\ {\em before} the other emission.}
\label{soft}
\end{figure}

However, many jet properties are sensitive to soft gluon emission, such
as the sub-jet distributions.  In this case, it seems at first sight
like no probabilistic algorithm could possibly describe the full matrix
element.  This is because the matrix-element amplitude for any given
final-state configuration is the sum of terms in which the gluon is
emitted by each external leg, as illustrated in Fig.~\ref{soft}.
Unlike the collinear case, all of these contribute to the leading
logarithm and when taking the square of the sum the interference terms
can be both positive and negative.  Therefore it looks like we have to
abandon the idea of describing the production of soft gluons in terms of
independent probabilistic $1\to2$ splittings.

However the fact that radiation from all the different emitters is
coherent, a simple consequence of gauge invariance, comes to the rescue.
It means that, after averaging over the relative azimuths of different
emissions\footnote{In fact azimuth-dependent terms that average to zero
  can also be included in the parton shower framework \cite{ian} and are
  automatically incorporated in the colour dipole cascade
  model \cite{cdm}.}, the emission of any given soft gluon can be
attributed to a single coloured line somewhere in the diagram, either
external, or internal {\em but imagined to be on-shell} \cite{bible}.
The fact that the emission probability from an internal line is what it
would be if it was on-shell results in the famous angular ordering
condition \cite{mw1}, namely that large angle emission should be treated
as occuring earlier than smaller angle emission.  This is the basis of
{\em coherent} parton shower algorithms.

It is important to note that this is not the same as using a standard
virtuality-ordered algorithm and disallowing disordered emission,
although claims are often made to the contrary.  For example, in the
configuration of Fig.~\ref{soft}, the total gluon emission probability
is non-zero and proportional to~$C_F,$ the colour charge of a quark,
whereas the veto algorithm would simply disallow all soft large-angle
emission from this system.

Many of the most important next-to-leading logarithmic contributions can
be incorporated by simple modifications to the leading logarithmic
coherence-improved algorithm.  A well-known example is the argument of
the running coupling~--- large sub-leading corrections can be summed to
all orders simply by using transverse momentum.

Emission from initial-state coloured partons can also be treated in the
same way.  Collinear emission is responsible for the scaling violations
in the parton distribution functions.  This means that the input set of
p.d.f.s can be used to guide the collinear emission and ensure that the
distributions produced are the same as those used in theoretical
calculations.  This allows a `backwards' evolution algorithm to be
used \cite{nrojbrot} with evolution starting at the hard interaction and
working downwards in scale back towards the incoming hadron.  Partons
produced by initial-state radiation subsequently undergo final-state
showering.  The coherence of radiation from different sources can again
be incorporated by choice of the evolution scale, at least for not too
small momentum fractions\footnote{Even at asymptotically small $x,$ it
  is possible to construct a probabilistic parton shower
  algorithm \cite{smallx}, but at present this is only of the forward
  evolution type, making it vastly inefficient for most studies.
  However, the authors of~\cite{smallx} found very little discrepancy
  with their coherent parton shower algorithm \cite{mw2}, even at the
  very small $x$ values encountered in DIS at a LEP+LHC energy.}.  In
this case the appropriate variable is the product of the opening angle
and the emitter's energy \cite{mw2}.

\begin{figure}[b]
\centerline{\psfig{file=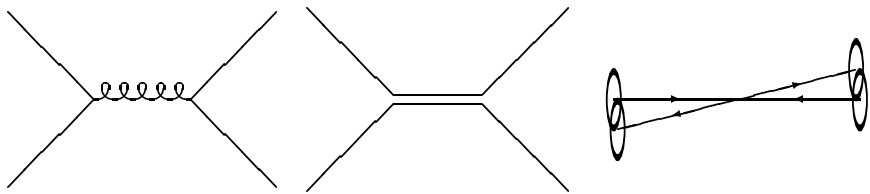,height=1.1in}}
\caption[]{Feynman (left), colour flow (centre) and cmf frame (right)
  diagrams of $\mathrm{q\bar{q}\to q'\bar{q}'}$ showing the radiation
  cones of each parton.  Most processes have many different colour
  flows.}
\label{colour}
\end{figure}

So far I have described how the parton shower evolves, but it also needs
a starting condition to evolve from.  This is particularly important for
{\em interjet} properties, as well as for setting the whole scale of the
subsequent evolution.  It is provided by the lowest order matrix element
for the process under consideration~--- jet production, prompt photon,
top quark production and decay, or whatever.  As shown in~\cite{keith},
the coherence of radiation from the various emitters plays an important
r\^ole here too.  Each hard process can be broken down into a number of
`colour flow' diagrams, which control the pattern of soft radiation in
that process.  To leading order in the number of colours, $N_c,$ gluons
can be considered as colour-anticolour pairs and a unique `colour
partner' can be assigned to each parton for each colour flow.  After
azimuthal averaging, radiation from each parton is limited to a cone
bounded by its colour connected partner, as illustrated for a
particularly simple case in Fig.~\ref{colour}.
It is important to note that although the radiation inside the cone
around each parton is modeled as coming from that parton, it is the
coherent sum of radiation from all emitters in the event {\em even the
  internal lines}, a point that I shall return to later.

Violations of this picture come from two main sources~---
colour-suppressed terms and semi-soft terms.  $1/N_c$ is not such a
small expansion parameter so one might expect non-leading colour terms
to be a very important correction.  However, they tend to be also
dynamically suppressed, since they are non-planar, and neglecting them
is generally a good approximation.  One can however find special corners
of phase-space in which no parton shower algorithm based on the
large-$N_c$ limit could be expected to be reliable \cite{crucis}.  The
colour-suppressed terms are generally negative and have not successfully
been incorporated into a probabilistic Monte Carlo picture.  The
semi-soft corrections arise simply from the fact that the emission cones
are derived in the limit of extremely soft emission that does not
disturb the kinematics of the event at all, while harder emission makes
the emitters recoil, disturbing their radiation patterns.  Both of these
effects mean that the initial conditions used in most implementations of
coherent parton showers are actually too strong, since they prevent
emission in regions where it is suppressed but not absent in the full
calculation (i.e.~outside the cones).

\clearpage
\subsubsection*{Monte Carlo Parton Showers}
Monte Carlo shower algorithms implement the probabilistic interpretation
in a very literal way, using a random number generator.  They give fully
exclusive distributions of all final-state properties with unit weight,
meaning that any given configuration occurs with the same frequency as
in nature.  The relevant jet properties can be directly measured from
the produced particles, exactly as in the experimental procedure.

The available models were thoroughly reviewed in \cite{argonne}, so I
only make a few brief comments about each.

\vspace{\topsep}\noindent{\bf ISAJET} \cite{isajet} implements a
virtuality-ordered collinear parton shower algorithm for both initial-
and final-state evolution.  It does not include any account of
coherence.  It is specific to hadron-hadron collisions, so its
parameters can be freely tuned to improve the fit to data.

\vspace{\topsep}\noindent{\bf PYTHIA} \cite{pythia} also implements a
virtuality-ordered collinear parton shower algorithm.  In the case of
final-state radiation, this is coherence-improved by disallowing
emission with disordered opening angles.  Coherence is now incorporated
into the initial conditions for the initial-state radiation (this is
called `PYTHIA+' below and is now the default), but not within the
evolution itself.  The parameters are those tuned to $\mathrm{e^+e^-}$
annihilation so the model is already well-constrained and predictive for
hadron-hadron collisions.

\vspace{\topsep}\noindent{\bf HERWIG} \cite{herwig} implements a complete
coherence-improved parton shower algorithm for both initial- and
final-state evolution, incorporating azimuthal correlations both within
and between jets.  The implementation is sufficiently precise that in
limited regions of phase-space it is reliable to next-to-leading
logarithmic accuracy, and HERWIG's $\Lambda$ parameter can be related to
$\Lambda_{\overline{\mathrm{MS}}}$ \cite{lambdamc}.  The parameters are
again tuned to $\mathrm{e^+e^-}$ and DIS data making the model highly
predictive.

\vspace{\topsep}\noindent{\bf ARIADNE} \cite{ariadne} is a new Monte
Carlo event generator that has only recently become available for
hadron-hadron collisions, although it has been very successful in
describing $\mathrm{e^+e^-}$ and DIS data.  It fully implements colour
coherence, but in a completely different way to that described above,
being based on the colour dipole cascade model \cite{cdm}.  It makes no
separation into initial- and final-state radiation, instead modeling
the emission from the whole system in a coherent way.  Its parameters
are also tuned to $\mathrm{e^+e^-}$ and DIS data.

\vspace{\topsep}

\begin{figure}
\vspace*{-2ex}
\centerline{\psfig{file=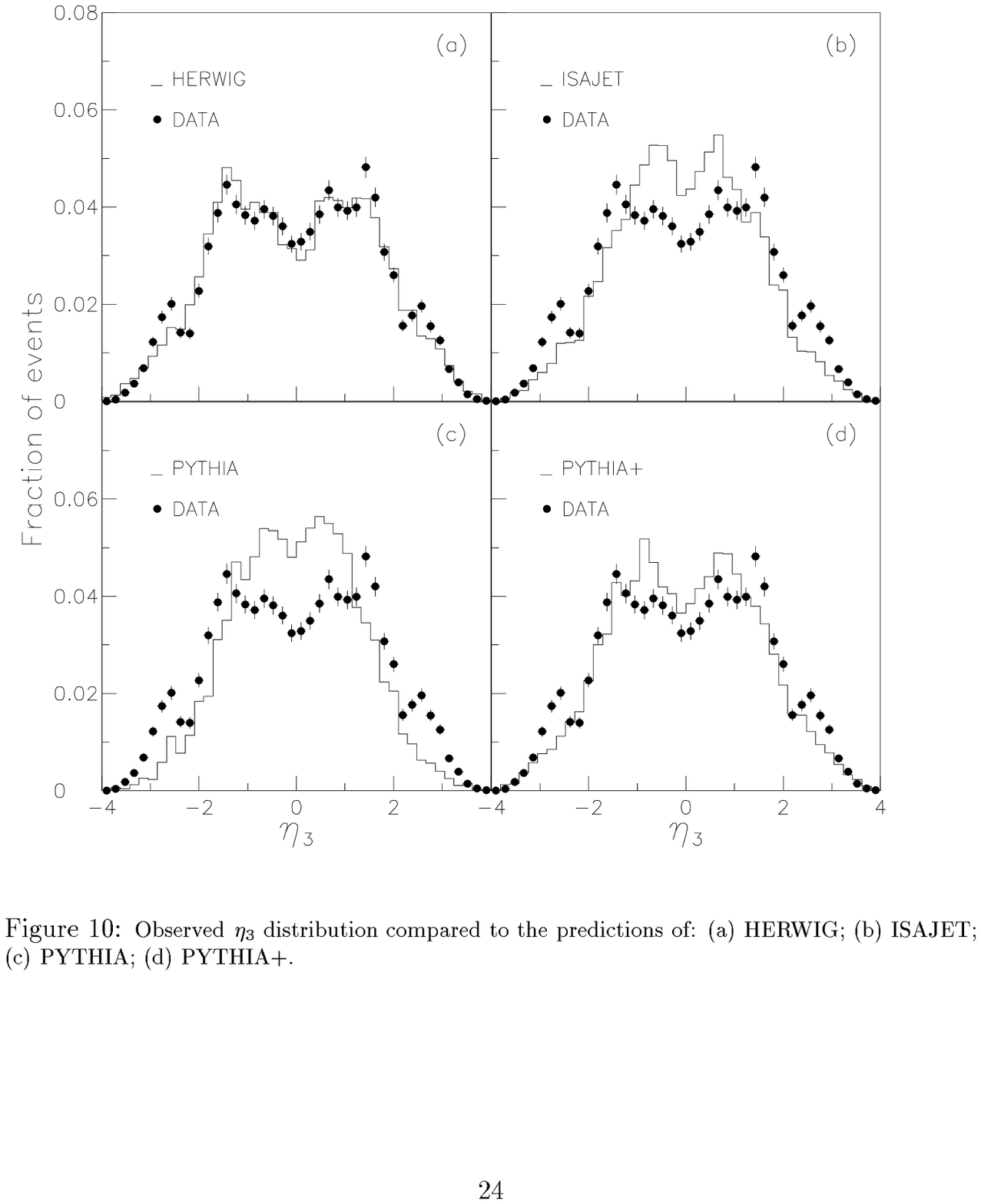,height=3in,clip=}}
\caption[]{The rapidity distribution of the third-hardest jet in
  multi-jet events, taken from~\cite{cdfco}.}
\vspace*{-1ex}
\label{CDFco}
\end{figure}

The issue of coherence has become paramount in describing
$\mathrm{e^+e^-}$ data, and models that do not implement it, at least
approximately, are completely ruled out (see for example~\cite{leprev}
and references therein).  They are also strongly disfavoured in DIS.
Data from the collider are now sufficiently precise that coherence
effects can also be studied there.  Fig.~\ref{CDFco} shows the results
of a recent CDF study \cite{cdfco} of the distribution of the softest jet
in three jet events.
Only the models incorporating coherence, HERWIG and the updated PYTHIA,
can account for the data.  It is possible that a similar interpretation
could be made of the sub-jet data presented in~\cite{rich}, although
further study is needed to confirm this.

\vspace*{-2ex}
\subsubsection*{Hard Parton Shower Emission}
\vspace*{-1ex}
By construction, coherence-improved shower algorithms reproduce the
exact matrix element in the soft and collinear limits.  Soft here means
relative to the hardest scale in the event~--- a jet of 50~GeV or so
could be considered soft in the context of top quark production for
example.  However, jet cuts always pull us away from the soft and
collinear regions since they require the emission to be resolvable.  For
example, the cross-section for `Mercedes' events, with three jets of
equal transverse momentum evenly spaced in azimuth, is not completely
negligible and since this is so far from any singular regions, we should
certainly worry how well parton shower algorithms will reproduce it.
There are two separate worries~--- whether there are phase-space regions
that the algorithm does not populate and whether it is a good enough
approximation in the regions is does populate.

The accuracy of fixed-order matrix elements is complementary to that of
parton showers~--- they are exact for hard well-separated jets (to
leading order in $\alpha_s$), but do not correctly account for multiple
emission effects, so become increasingly unreliable if the cutoffs are
made very small.  It is natural to hope that the two approaches can be
combined to yield a single algorithm that is uniformly reliable for hard
and soft emission, collinear and well-separated.  Several
phenomenological attempts have been made in the past, but these
generally suffer a variety of problems.  Firstly, it is essential that
if the phase-space region is divided into two parts then they are
smoothly matched at the boundary to prevent double-counting or
under-counting.  Ideally the boundary itself should be an arbitrary
parameter so that one can explicitly check that varying it does not
affect the output distributions.  It is worth noting that for this to be
true, it is incorrect to use the exact leading order matrix element, as
a form factor must be introduced even within the hard region.  Many
algorithms work by correcting the {\em first} emission to reproduce the
hard matrix element.  However, the choice of ordering variable is
scheme-dependent and so therefore is the definition of which emission is
first.  This problem is particularly severe in angular-ordered parton
shower algorithms, where there are often several soft large-angle
emissions before the hardest emission in the jet.  If one nevertheless
just corrects the first emission, one obtains a hard limit that is
dependent on the soft infrared cutoff of the algorithm, which is clearly
unphysical.

\begin{figure}[b]
\vspace*{-2ex}
\centerline{\psfig{file=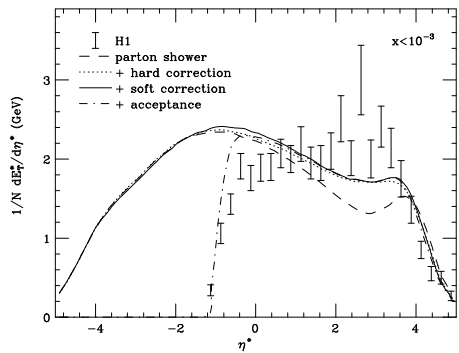,height=2in}}
\vspace*{-1ex}
\caption[]{The transverse momentum flow in the $\gamma^*\mathrm{p}$
  centre-of-mass frame of DIS events at small $x$, taken
  from~\cite{glasgow}, data from~\cite{h1}.  The first three curves are
  without detector acceptance and the dot-dash is the third after
  including the acceptance.}
\vspace*{-1ex}
\label{glasgow}
\end{figure}

These problems were discussed in \cite{wellhard}, where self-consistent
algorithms were proposed for correcting both deficiencies~--- filling
empty regions of phase-space and correcting the distributions of all
hard emission (not just the first) within the parton shower region.
There is no conceptual barrier to applying them to arbitrary hard
processes, such as top quark production and decay, and it would be
straightforward to do so in an algorithm specifically designed with this
in mind.  However, it has proved rather tricky to weave them into the
existing HERWIG algorithm and this has so far only been done for the
simplest processes, $\mathrm{e^+e^-\to q\bar{q}}$ and DIS,
$\mathrm{eq\to eq}$.  Results for the latter are shown in
Fig.~\ref{glasgow}\footnote{In fact almost perfect agreement with the
  data is obtained if the default parameters tuned to $\mathrm{e^+e^-}$
  annihilation are used.  This figure instead uses H1's tuned set for
  comparison with their paper.}.
The hard correction, i.e.\ the filling of empty phase-space regions is
an important correction, while the soft correction, within the
parton-shower phase-space makes very little difference.  The same is
also true in $\mathrm{e^+e^-}$.  This can be interpreted as meaning that
although HERWIG's parton shower algorithm does not fill the whole of
phase-space owing to its over-strict implementation of the
angular-ordering initial condition, it is a good approximation in the
regions it does fill.

It seems plausible that these corrections would fix the problems for top
quark production and decay events reported in~\cite{oss} and in Stephen
Parke's talk \cite{stephen}, but we have not yet been able to explicitly
check this.

\subsubsection*{Azimuthal Decorrelation}
In the last couple of years there have been several proposed tests of
`non-DGLAP' evolution, i.e.~quantities for which the collinear evolution
described above is insufficient, as discussed in the talk by Vittorio
del Duca~\cite{vittorio}.  Although some of these are certainly
sensitive to new small-$x$ dynamics, many can be reliably predicted
using coherence-improved evolution.

\begin{figure}[b]
\centerline{\psfig{file=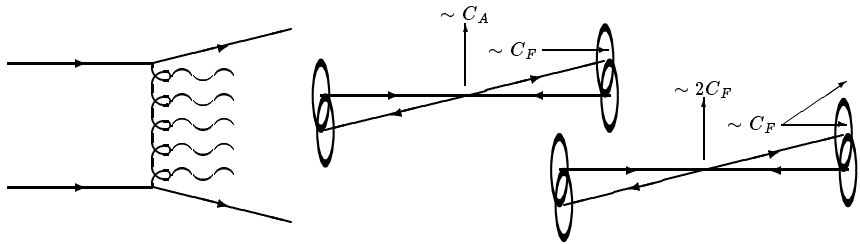,height=1.3in}}
\caption[]{BFKL ladder (left), full emission pattern (centre) and
  leading-$N_c$ emission pattern (right) for small angle $\mathrm{qq'}$
  scattering, contributing to the azimuthal decorrelation of the
  resulting jets.  In the full case, emission from each quark is
  confined to its small forward cone and there is emission from the
  internal gluon, while in the leading-$N_c$ case each quark emits
  everywhere except its small backward cone and there is no other
  emission.}
\label{bfkl?}
\end{figure}

One such example is the decorrelation in azimuth of jet pairs as the
rapidity interval between them increases.  In the BFKL framework, this
is attributed to emission from the $t$-channel exchanged gluon, as
illustrated in Fig.~\ref{bfkl?} for a simple example, $\mathrm{qq'\to
  qq'},$ chosen because it has a single colour flow.
The full emission pattern of this system consists of emission
proportional to~$C_F$ inside small cones of opening angle~$\theta_s,$
the scattering angle and proportional to~$C_A$ in the remainder of the
solid angle.  However, a coherence-improved parton shower would describe
this situation as two colour lines, each of which have been scattered
through almost~$180^\circ$.  Thus each quark emits proportional to~$C_F$
into a `cone' of opening angle $180^\circ\!-\theta_s$ and, apart from
the difference between $C_A$ and $2C_F,$ a $1/N_c^2$ correction, it
reproduces the full emission pattern.  Thus to leading order in the
number of colours, coherence-improved parton showers include soft
emission from this internal line and should reproduce the full QCD
result for azimuthal decorrelations.

\subsubsection*{Analytical Parton Showers}
\begin{figure}[b]
\vspace*{-2ex}
\centerline{\psfig{file=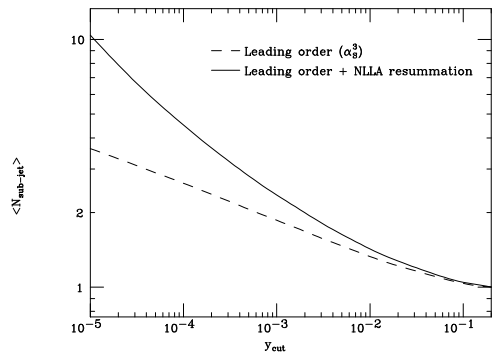,height=2in}}
\caption[]{The average number of sub-jets in a central 100~GeV jet,
  taken from~\cite{sub-jet}.}
\label{sub-jet}
\vspace*{-1ex}
\end{figure}

The probabilistic parton shower framework described above can also be
used for analytical calculations.  In this case, one sets up a
probabilistic evolution equation for how the quantity of interest, for
example the distribution of sub-jets, changes as a function of either
the jet $p_t$ or resolution scale.  This gives integro-differential
equations that, with appropriate boundary conditions, can be solved
analytically.  However, this can only be done for experimental
observables in which the phase-space for multiple resolved emission
factorizes.  In the case of sub-jets this essentially requires that the
resolution variable be of $k_\perp$ type.  So far, only one calculation
of this type has been made for hadron-hadron collisions (although there
are several for $\mathrm{e^+e^-}$ and DIS), for the average number of
sub-jets resolved in an inclusive jet \cite{sub-jet}.  The result is
shown in Fig.~\ref{sub-jet} and is compared to D0 data in~\cite{rich}.
The increasing importance of the all-orders resummation of
next-to-leading logarithmic terms at small $y_{\mathrm{cut}}$ can be
clearly seen.

\subsection*{Hadronization}
\vspace*{-1ex}
The process by which coloured partons are confined into hadrons is not
understood at a fundamental level at present, so phenomenological models
must be used.  The dynamics of the parton shower are {\em preconfining},
meaning that partons tend to end up close, in both phase-space and
real-space, to their colour-connected partners.  This suggests that
hadronization is a fairly local process, taking place in the spatial
regions between (but near to) the final-state partons.  The string and
cluster models used in PYTHIA and\linebreak[3] HERWIG respectively both
implement this idea and both give good fits to $\mathrm{e^+e^-}$ data.
Like their parton shower algorithms, the model parameters are strongly
constrained by $\mathrm{e^+e^-}$ data, making them highly predictive for
hadron-hadron collisions.  The independent fragmentation model, in which
single partons decay to hadrons according to longitudinal phase-space
with no account of colour connections, is already ruled out by
$\mathrm{e^+e^-}$ data.

Owing to lack of time in the talk and space in the proceedings, I can do
no more than mention a new area that is sure to grow in the near future,
the analytical calculation of hadronization corrections.  Several
different approaches appear to be converging on the result that
perturbation theory, suitably modified, is more powerful than we thought
and may even be capable of predicting hadron-level cross-sections.  The
introduction of an effective running coupling that is integrable at low
momenta \cite{dokweb} alone may be sufficient to calculate the dominant
corrections to infrared-finite quantities, although the factorization
this implies \cite{akhoury} has been questioned in renormalon-based
approaches \cite{us,them}.  More details can also be found in George
Sterman's talk \cite{george}.

\vspace*{-1ex}
\section*{Conclusion}
\vspace*{-1ex}
As I have stressed throughout the talk, jets are not fundamental objects
in QCD, but are artificial event properties defined by hand.  Different
definitions have good and bad points and the results will be a function
of these definitions.  We need reliable methods to calculate these event
properties from the fundamental objects of perturbative QCD, quarks and
gluons.

Cone-type jet algorithms are simple in principle, but turn out to be
complicated in practice, owing to jet overlap problems.  Cluster-type
algorithms are more complicated to define, but turn out to be simple in
practice, since exactly the same algorithm can be used for one or two
partons as on the hadron-level or detector-level final state.  The
$k_\perp$ cluster algorithm has a variety of practical and theoretical
advantages both for jet physics and for event reconstruction.

For jet properties, very few analytical calculations are available,
particularly in the important region of small resolution parameters, but
this is expected to improve in the future.  Modern coherence-improved
Monte Carlo models are very sophisticated implementations of
perturbative QCD plus very constrained models of hadronization and are
highly predictive for hadron-hadron collisions.  However, there are
areas in which they need to be improved, most notably in matching parton
showers with exact matrix elements to improve the description of very
hard emission.

It is clear that many areas outside the standard QCD arena are becoming
increasingly reliant on jet physics.  For example the error on the top
mass will soon become dominated by the treatment of gluon radiation and
it will become absolutely essential to consider the problems discussed
in this talk.  Namely, how can the present jet definition and modeling
be improved upon?

\end{document}